# Inflatable 'Evergreen' Polar Zone Dome (EPZD) Settlements

**Alexander A. Bolonkin**[*]
Space scientist
1310 Avenue R, Suite 6-F
Brooklyn, New York 11229, USA
aBolonkin@juno.com
(718) 339-4563

**Richard B. Cathcart**
Geographos
1300 West Olive Avenue, Suite M
Burbank, California 91506, USA
rbcathcart@charter.net
(818) 953-9113

**ABSTRACT**

Sustaining human life at the Earth's antipodal Polar Regions is very difficult especially during Winter when water-freezing air temperature, blizzards and "whiteouts" make normal human existence dangerous. To counter these environmental stresses, we offer the innovative artificial "Evergreen" Polar Zone Dome (EPZD), an inflated half-hemisphere with interiors continuously providing a Mediterranean Sea-like climate. The "Evergreen" EPZD structural theory is developed, substantiated by key computations that show it is possible for current building technology to construct and heat large enclosed volumes inexpensively. Specifically, a satisfactory result is reached by using sunlight reflectors and a special double thin film, which concentrates all available solar energy inside the EPZD while, at the same time markedly decreasing the heat loss to exterior Polar Region air. Someday a similar, but remarkably more technological, EPZD design may be employed at proposed Moon and Mars settlements.

## I. INTRODUCTION

Particularly during Winter, the Earth's two Polar Regions provide only a meager and uncomfortable life-style for humans, featuring very low ambient air temperature, strong wind, and seasonal darkness. Starting from Cape Artichesky (Russia), Mike Horn and Borge Ousland completed, during January to March 2006, the first known over-pack ice trek to the planet's geographic North Pole during darkness! More persons are likely to undertake such journeys near the Arctic Ocean's coast when permanent dwellings are situated closer to the North Pole. The purpose of our report is to make possible economical new cities at both Polar Regions of the Earth.

Economists allege that the mean 2006 USA Dollar value of Polar Region land territory is generally low compared to the world total of ~$250,000/km$^2$. For example: Antarctica ~$40/km$^2$, Greenland ~$650/km$^2$, Canada ~$77,000/km$^2$ and Russia ~$106,000/km$^2$. However, world economic productivity data show that the 2006 USA dollar output per capita in the Earth-biosphere is greatest in Polar Regions; cold regions have output per capita that is approximately 10-12 times that of the Earth's Tropic Zones! The distance north or south from the planet's Equator is amongst the most significant measured environmental variables underlying the differences expressed in per capita USA dollar output by country-ecosystem, but this is probably explained by the overall global pattern of human settlement, which tends to influence social institutions and their supportive technologies.[1]

In other words, if persons living at the Polar Regions were made more comfortable than now, it is very probable that the economic value per square kilometer of territory situated in those two geographical regions would decrease slightly since only very poor persons (nomadic natives in the Northern Hemisphere) and highly paid persons (technicians and scientists[2]) dwell fulltime in the Earth's Polar Regions nowadays! Non-nomadic people currently work in these uncomfortably cold and seasonally dark climate settlements only because there are known mineral, natural gas and petroleum deposits to be mined, along with seasonal and non-seasonal hydroelectric facilities[3] to be efficiently operated for the benefit of large populations living in warmer climates away from primary production places. There is every reason to think that valuable Arctic and Antarctica resources remain undiscovered, awaiting future exploration and industrial exploitation. The Arctic alone has proven discovered oil and natural gas deposits equal to 40% of Saudi Arabia's total reserves.[4] Many people worldwide, especially in the Temperate Zones, muse on the possibility of humans someday inhabiting orbiting Space Settlements and Moon Bases or a terraformed Mars but few seem to contemplate an increased use of ~25% of Earth's surface—the Polar Regions.[5] Antarctica is being investigated for its economic potential[6] and already the Antarctic Circumpolar Current has been affected by global civilization.[7]

---
[*] Corresponding Author.

## II. ARCTIC

Geoscience has generally substantiated that the Arctic is warming and eventually it may reach a seasonally "ice-free" state caused by the absence of new sea-ice due to non-formation as well as summertime excess sublimation of glaciers on land during the 21$^{st}$ Century.[8] Climatological feedback loops, which effect change, are the interplay between sea/land ice, North Atlantic Ocean currents—especially the Gulf Stream and the north-flowing current in the Bering Strait—and the annual amounts of precipitation and evaporation in the Arctic. Reduced sea-ice extent and thickness in the Arctic Ocean would promote regular summertime commercial shipping, and present new opportunities for offshore oil and gas extraction. A Northern Sea Route paralleling the Siberian coastline would be ~40% shorter than the current Europe-Asia Route which requires a passage through the Suez Canal.[9] In addition, new macroprojects—opportunistic hydroelectric power development of diminishing glaciers and a permanent tunnel or bridge[15] across the Bering Strait—may attract new settlers to the Arctic. There is also the possibility, as we explore here, for Arctic Zone greenhouses under inflated membrane half-hemispheres producing fresh fruits and vegetables for workers on such macroprojects and to luxuriously house at low-cost workers seeking to maintain the present-day natural stock of Arctic Ocean sea-ice by construction of artificial ice islands.[10]

## III. 'EVERGREEN' INFLATED DOMES

Possibly the first true architectural attempt at constructing effective artificial life-support systems in climatically harsh regions in the Earth-biosphere was the building of greenhouses. Extensive commercial greenhouses in The Netherlands—and even outer space[11]—are maintained nearly automatically by heating, cooling, irrigation, nutrition and plant disease management equipment. The "Climatron" greenhouse was finished in 1959 at the Missouri Botanical Garden in St. Louis, USA, while the world's most voluminous greenhouses, the Eden Project, were completed in Cornwall, UK during the early 21$^{st}$ Century.[12] All people share commonalities in their responses to natural environmental stresses that are stimulated by extremely cold air, snowstorms and strong wind. In the Arctic and Antarctica, life-threatening "whiteout" snowstorms inflict somewhat the same personal visual discomfort and disorientation as cosmonauts/astronauts experience during their space-walks—that of being adrift in featureless outer space! With special clothing and shelters, humans can adapt to these Polar Regions successfully. Medical researchers have asserted that "…cold-related deaths are far more numerous than heat-related deaths in the United States, Europe, and almost all countries outside of the tropics, and almost all of them are due to common illnesses that are increased by cold."[13] Incontrovertibly, living near the planet's poles is stressful and operationally difficult, even when tempered by strong conventional buildings such as those at the Earth's South Pole where the Ozone Hole causes a UVB radiation hazard that cannot be ignored because it helps cause sunburn (erythema) and snow blindness (photokeratitis); the Arctic also has a UVB radiation hazard. The morale of personnel—a difficult factor to measure—during wintertime when little daylight and little contrast between land, sea and sky predominates can cause monotony, a negative influence on personnel activity and efficiency. Essentially, any EPZD becomes the total environment of its inhabitants, so proper internal temperature control and soundproofing are vital. Relative humidity inside the EPZD ought to be fixed at 30-40% for human comfort and health reasons and to insure that static electricity does not become a problem affecting safety in the EPZD.

The first "Evergreen"-type dwelling hemisphere design "City in the Arctic" was commissioned by Germany's Farbwerke Hoechst AG in 1970.[14] "City in the Arctic" was a pneumatically stabilized climate-regulating transparent membrane half-sphere shell with a diameter 2,000 to 2,200 m and a height of about 240 m intended to comfortably enable 15,000 to 45,000 workers. The contemplated membrane was to be reinforced and supported by a net of intersecting, braided polyester fiber cables. Better 21$^{st}$ Century materials are available that would improve the formidable performance characteristics of "City in the Arctic".[15] Founded in 1956, Birdair, Inc. in the USA, introduced its Sheerfill™ in 1970 and this constantly improved material offers translucencies of ~25%. The energy impact of Sheerfill, and other nearly non-combustible architectural technical textiles[16], is a function of the tradeoff between decreased lighting needs and increased heating costs; a dynamic air-supported membrane building normally costs only about 33% of a building assembled with ordinary materials.[17] (More technologies will be revealed at The Twelfth International Workshop on the Design and Practical Realization of Architectural Membrane Structures, "Textile Roofs 2007", held 7-9 June 2007 in Berlin, Germany.) "City in the Arctic" was never built because it was merely an architectural speculation, even less substantial than the architectural speculations of Sotirios Kotoulas in ***Space Out*** (Springer, 2006).

Our macro-engineering concept of inexpensive-to-construct and operate "Evergreen" inflated Earth-surface domes is supported by computations, making our macroproject speculation more than a daydream. However, we lack access to a low-speed laboratory wind tunnel and that inhibits our apprehension of wall interference and the effect of some layout details, such as open or closed doors facing the EPZD exterior. Innovations are needed, and wanted, to realize such structures in the Polar Regions of our unique but continuously changing planet.

### IV. DESCRIPTION AND INNOVATIONS

Our design for an Arctic people-housing "Evergreen" dome is presented in Fig. 1, which includes the thin inflated film dome. Air-supported construction derives from the balloon principle to shape a building; the air pressure inside the building exceeds the external air pressure to support the roof. Sunlight can penetrate special roofing materials, making the interiors brighter than others types of constructed buildings. The EPZD innovations are listed here: (1) the construction is air-inflatable; (2) each dome is fabricated with very thin, transparent film (thickness is 0.1 to 0.2 mm) without rigid supports; (3) the enclosing film is a two-layered element with air between the layers to provide insulation; (4) the construction form is that of a hemisphere, or in the instance of a roadway/railway a half-tube, and part of it has a thin aluminum layer about 1 μ or less that functions as the gigantic collector of solar incident solar radiation (heat). Surplus heat collected may be used to generate electricity or furnish mechanical energy; and (5) the dome is equipped with sunlight controlling louvers [AKA, "jalousie", a blind or shutter having adjustable slats to regulate the passage of air and sunlight] with one side thinly coated with reflective polished aluminum of about 1 μ. Real-time control of the sunlight's entrance into the dome and nighttime heat's exit is governed by the shingle-like louvers.

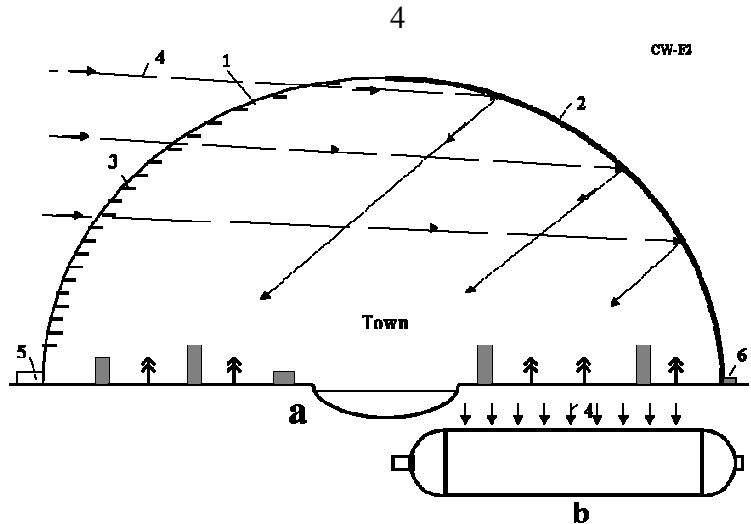

**Fig.1.** Artificial inflatable dome for Earth's Northern Hemisphere cold regions. Notations: (a) cross-section area of suggested biosphere; (b) top view of cylindrical biosphere, 1 - transparence thin double film ("textiles"); 2 - reflected cover of half-hemisphere; 3 - control louvers (jalousie); 4 - solar beams (light); 5 - enter; 6 - air pump (ventilator).

Fig. 1 illustrates the thin transparent dome cover we envision. The hemispherical inflated textile shell—technical "textiles" can be woven or non-woven (films)—embodies the EPZD innovations listed: (1) the film is very thin, approximately 0.1 to 0.2 mm. A film this thin has never before been used in a major building; (2) the film has two strong nets, with a mesh of about 0.1 × 0.1 m and $a = 1 \times 1$ m, the threads are about 0.3 mm for a small mesh and about 1 mm for a big mesh. The net prevents the watertight and airtight film covering from being damaged by vibration; (3) the film incorporates a tiny electrically conductive wire net with a mesh about 0.01 x 0.01 m and a line width of about 100 μ and a thickness near 1μ. The wire net can inform the "Evergreen" dome supervisors concerning the place and size of film damage (tears, rips, etc.); (4) the film is twin-layered with the gap — $c = 1$ m and $b = 2$ m—between covering's layers. This multi-layered covering is the main means for heat insulation and puncture of one of the layers wont cause a loss of shape because the film's second layer is unaffected by holing; (5) the airspace in the dome's covering can be partitioned, either hermetically or not; (6) the units #5 of the cover is furnished with a heat tube #6 that can spray warmed anti-freeze liquid onto the EPZD's exterior, thus eliminating snow and ice buildup; and (7) part of the covering has a very thin shiny aluminum coating that is about 1μ for reflection of incoming solar radiation.

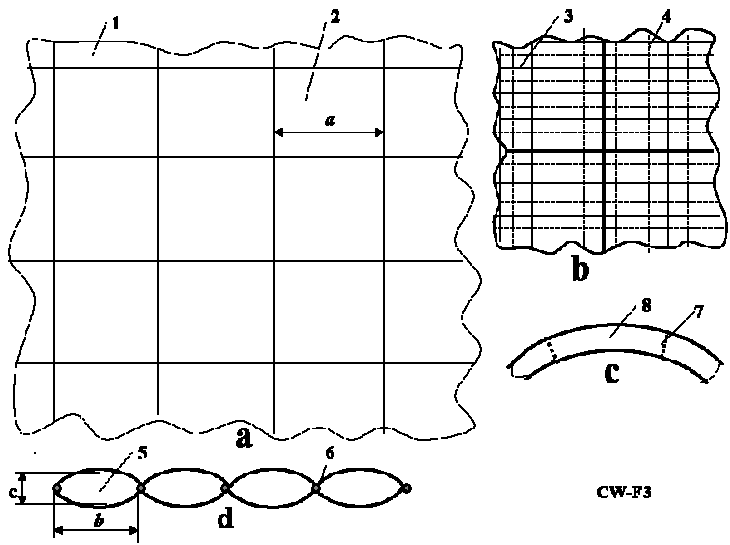

**Fig.2.** Design of membrane cover. Notations: (a) Big fragment of cover; (b) Small fragment of cover; (c) Cross-section of cover; (d) Longitudinal cross-section of cover; 1 - cover; 2 -mesh; 3 - small mesh; 4 - thin electric net; 5 - sell of cover; 6 - tubes; 7 - film partition (non hermetic); 8 - perpendicular cross-section area.

Fig. 3 illustrates a lightweight, possibly portable house using the same essential construction materials as the dwelling/workplace shown in Fig. 1.

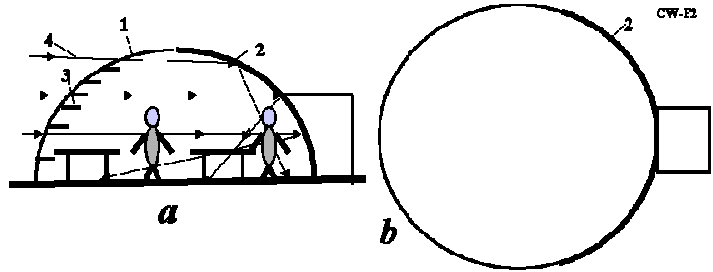

**Fig. 3.** Inflatable film house for cold climate regions. Notation: (a) Cross-section area; (b) Top view. The other notations are same with fig.1.

## V. THEORY AND COMPUTATIONS EPZD

As wind flows over and around a fully exposed, nearly completely sealed inflated dome, the weather affecting the external film on the windward side must endure positive air pressures as the wind stagnates. Simultaneously, low air pressure eddies will be present on the leeward side of the dome. In other words, air pressure gradients caused by air density differences on different parts of the sheltering dome's envelope is characterized as the "buoyancy effect". The buoyancy effect will be greatest during the coldest weather when the EPZD is heated and the temperature difference between its interior and exterior are greatest. In extremely cold climates, such as the Arctic and Antarctica, the buoyancy effect tends to dominate dome pressurization, causing the EPZD to require reliable anchoring.

Our basic computed equations, below, are derived from a Russian-language textbook.[18] Solar radiation impinging the orbiting Earth is approximately 1400 W/m$^2$. The average Earth reflection by clouds and the sub-aerial surfaces (water, ice and land) is about 0.3. The Earth-atmosphere adsorbs about 0.2 of the Sun's radiation. That means about $q_0$ = 700 W/m$^2$s of solar energy (heat) reaches our planet's surface at the Equator. The solar spectrum is graphed in Fig. 4.

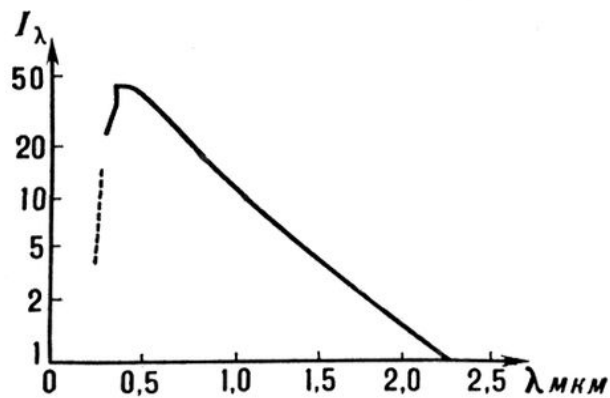

**Fig.4.** Spectrum of solar radiation. Visible light is 0,4 - 0,8 $\mu$.

The visible part of the Sun's spectrum is only $\lambda$ = 0.4 to 0.8 $\mu$. Any warm body emits radiation. The emission wavelength depends on the body's temperature. The wavelength of the maximum intensity (see Fig. 4) is governed by the black-body law originated by Max Planck (1858-1947):

$$\lambda_m = \frac{2.9}{T}, \quad [mm], \qquad (1)$$

where $T$ is body temperature, $^{\circ}$K. For example, if a body has an ideal temperature 20 $^{\circ}$C ($T$ = 293 $^{\circ}$K), the wavelength is $\lambda_m$ = 9.9 $\mu$.

The energy emitted by a body may be computed by employment of the Josef Stefan-Ludwig Boltzmann law.

$$E = \varepsilon \sigma_s T^4, \quad [W/m^2], \qquad (2)$$

where $\varepsilon$ is coefficient of body blackness ($\varepsilon$ =0.03 ÷ 0.99 for real bodies), $\sigma_s$ = 5.67×10$^{-8}$ [W/m$^2 \cdot$K] Stefan-Boltzmann constant. For example, the absolute black-body ($\varepsilon$ = 1) emits (at $T$ = 293 $^0$C) the energy $E$ = 418 W/m$^2$.

Amount of the maximum solar heat flow at 1 m$^2$ per 1 second of Earth surface is

$$q = q_o \cos(\varphi \pm \theta) \quad [W/m^2], \qquad (3)$$

where $\varphi$ is Earth longevity, $\theta$ is angle between projection of Earth polar axis to the plate which is perpendicular to the ecliptic plate and contains the line Sun-Earth and the perpendicular to ecliptic plate. The sign "+" signifies Summer and the "-" signifies Winter, $q_o \approx$ 700 W/m$^2$ is the annual average solar heat flow to Earth at equator corrected for Earth reflectance.

This angle is changed during a year and may be estimated for the Arctic by the following the first approximation equation:

$$\theta = \theta_m \cos \omega, \quad \text{where} \quad \omega = 2\pi \frac{N}{364}, \qquad (4)$$

where $\theta_m$ is maximum $\theta$, $|\theta_m|$ = 23.5$^{\circ}$ =0.41 radian; $N$ is number of day in a year. The computations for Summer and Winter are presented in fig.5.

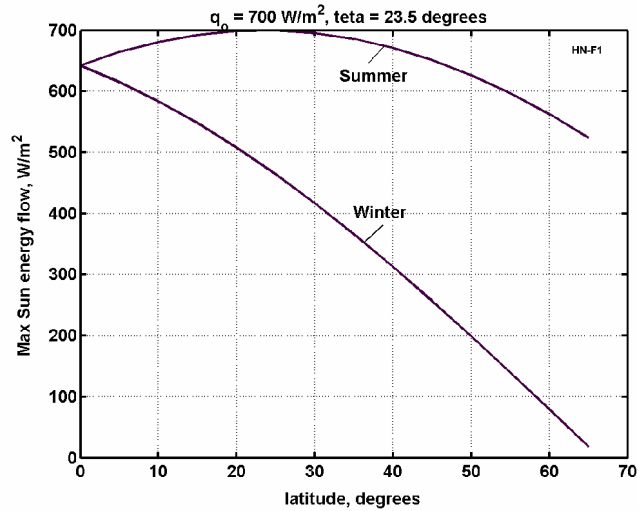

**Fig.5.** Maximum Sun radiation flow at Earth surface as function of Earth latitude and season.

The heat flow for a hemisphere having reflector (fig.1) at noon may be computed by equation

$$q = c_1 q_0 [\cos(\varphi - \theta) + S \sin(\varphi + \theta)], \qquad (5)$$

where $S$ is fraction (relative) area of reflector to service area of "Evergreen" dome. For reflector of Fig.1 $S$ = 0.5; $c_1$ is film transparency coefficient ($c_1 \approx$ 0.9 - 0.95).

The daily average solar irradiation (energy) is calculated by equation

$$Q = 86400 c q t, \quad \text{where} \quad t = 0.5(1 + \tan \varphi \tan \theta), \quad |\tan \varphi \tan \theta| \leq 1, \qquad (6)$$

where $c$ is daily average heat flow coefficient, $c \approx 0.5$; $t$ is relative daylight time, $86400 = 24 \times 60 \times 60$ is number of seconds in a day.

The computation for relative daily light period is presented in Fig. 6.

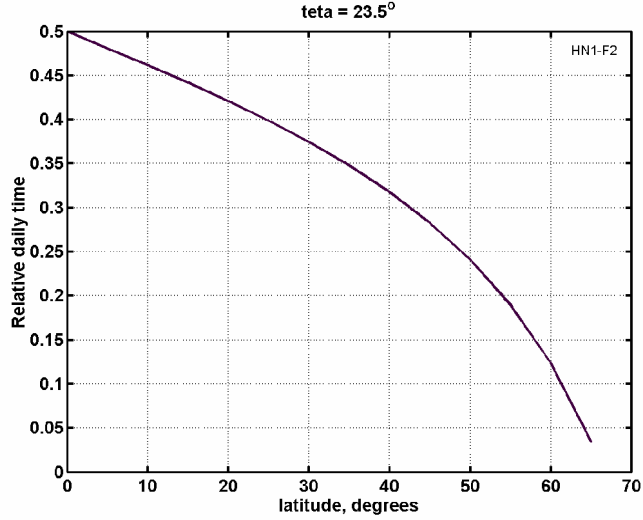

**Fig.6**. Relative daily light time relative to Earth latitude.

The heat loss flow per 1 m² of dome film cover by convection and heat conduction is (see [13]):

$$q = k(t_1 - t_2), \quad \text{where} \quad k = \frac{1}{1/\alpha_1 + \sum_i \delta_i / \lambda_i + 1/\alpha_2}, \qquad (7)$$

where $k$ is heat transfer coefficient, W/m²·K; $t_{1,2}$ are temperatures of the inter and outer multi-layers of the heat insulators, °C; $\alpha_{1,2}$ are convention coefficients of the inter and outer multi-layers of heat insulators ($\alpha = 30 \div 100$), W/m²K; $\delta_i$ are thickness of insulator layers; $\lambda_i$ are coefficients of heat transfer of insulator layers (see Table 1), m; $t_{1,2}$ are temperatures of initial and final layers °C.

The radiation heat flow per 1 m²s of the service area computed by equations (2):

$$q = C_r \left[ \left(\frac{T_1}{100}\right)^4 - \left(\frac{T_2}{100}\right)^4 \right], \quad \text{where} \quad C_r = \frac{c_s}{1/\varepsilon_1 + 1/\varepsilon_2 - 1}, \quad c_s = 5.67 \ [\text{W/m}^2\text{K}^4], \qquad (8)$$

where $C_r$ is general radiation coefficient, $\varepsilon$ are black body rate (Emittance) of plates (see Table 2); $T$ is temperatures of plates, °K.

The radiation flow across a set of the heat reflector plates is computed by equation

$$q = 0.5 \frac{C'_r}{C_r} q_r, \qquad (9)$$

where $C'_r$ is computed by equation (8) between plate and reflector.

The data of some construction materials is found in Table 1, 2.

**Table 1**. [13], p.331. Heat Transfer.

| Material | Density, kg/m³ | Thermal conductivity, $\lambda$, W/m·°C | Heat capacity, kJ/kg·°C |
|---|---|---|---|
| Concrete | 2300 | 1.279 | 1.13 |
| Baked brick | 1800 | 0.758 | 0.879 |
| Ice | 920 | 2.25 | 2.26 |
| Snow | 560 | 0.465 | 2.09 |
| Glass | 2500 | 0.744 | 0.67 |
| Steel | 7900 | 45 | 0.461 |

| | | 8 | | |
|---|---|---|---|---|
| Air | 1.225 | 0.0244 | | 1 |

---

As the reader will see, the air layer is the best heat insulator. We do not limit its thickness $\delta$.

**Table 2**. [13], p. 465. Emittance

---

| Material | Emittance, $\varepsilon$ | Material | Emittance, $\varepsilon$ | Material | Emittance, $\varepsilon$ |
|---|---|---|---|---|---|
| Bright Aluminum $t = 50 \div 500\ ^{\circ}C$ | 0.04 - 0.06 | Baked brick $t = 20\ ^{\circ}C$ | 0.88 - 0.93 | Glass $t = 20 \div 100\ ^{\circ}C$ | 0.91 - 0.94 |

As the reader will notice, the shiny aluminum louver coating is excellent mean jalousie against radiation losses from the dome.

The general radiation heat $Q$ computes by equation [6]. Equations [1] – [9] allow computation of the heat balance and comparison of incoming heat (gain) and outgoing heat (loss).
The computations of heat balance of a dome of any size in the coldest wintertime of the Polar Regions are presented in Fig. 7.

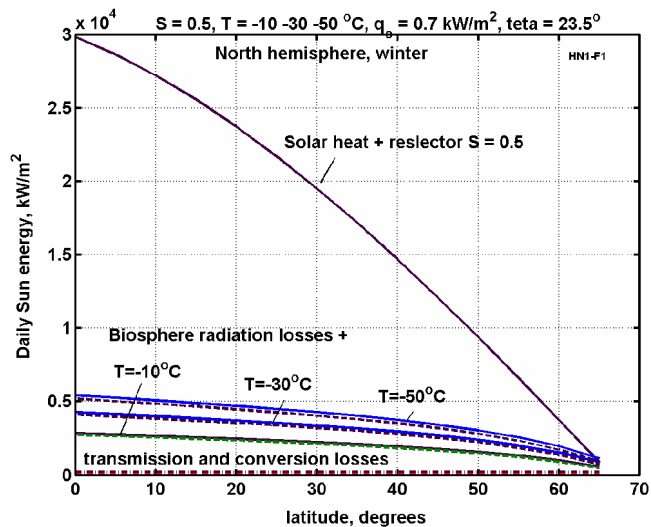

**Fig. 7.** Daily heat balance through 1 m² of EPZD during coldest winter day versus Earth's latitude (North hemisphere example). Data used for computations (see Eq. (1) - (9)): temperature inside of dome is $t_1 = +20\ ^{\circ}C$, outside are $t_2 = -10, -30, -50\ ^{\circ}C$; reflectivity coefficient of mirror is $c_2 = 0.9$; coefficient transparency of film is $c_1 = 0.9$; convectively coefficients are $\alpha_1 = \alpha_2 = 30$; thickness of film layers are $\delta_1 = \delta_2 = 0.0001$ m; thickness of air layer is $\delta = 1$ m; coefficient of film heat transfer is $\lambda_1 = \lambda_3 = 0.75$, for air $\lambda_2 = 0.0244$; ratio of cover blackness $\varepsilon_1 = \varepsilon_3 = 0.9$, for louvers $\varepsilon_2 = 0.05$.

The heat from combusted fuel is found by equation
$$Q = c_t m / \eta , \qquad (10)$$

where $c_t$ is heat rate of fuel [J/kg]; $c_t = 40$ MJ/kg for liquid oil fuel; $m$ is fuel mass, kg; $\eta$ is efficiency of heater, $\eta = 0.5 - 0.8$.
The thickness of the dome envelope, its sheltering shell of film, is computed by formulas (from equation for tensile strength):

$$\delta_1 = \frac{Rp}{2\sigma}, \quad \delta_2 = \frac{Rp}{\sigma}, \tag{11}$$

where $\delta_1$ is the film thickness for a spherical dome, m; $\delta_2$ is the film thickness for a cylindrical dome, m; $R$ is radius of dome, m; $p$ is additional pressure into the dome, N/m$^2$; $\sigma$ is safety tensile stress of film, N/m$^2$.

For example, compute the film thickness for dome having radius $R$ =100 m, additional air pressure $p$ = 0.01 atm ($p$ = 1000 N/m$^2$), safety tensile stress $\sigma$ = 50 kg/mm$^2$ ($\sigma$ = 5×10$^8$ N/m$^2$), cylindrical dome.

$$\delta = \frac{100 \times 1000}{5 \times 10^8} = 0.0002\, m = 0.2\, mm \tag{12}$$

The dynamic pressure from wind is

$$p_w = \frac{\rho V^2}{2}, \tag{13}$$

where $\rho$ = 1.225 kg/m$^3$ is air density; $V$ is wind speed, m/s.

For example, a storm wind with speed $V$ = 20 m/s, standard air density is $\rho$ = 1.225 kg/m$^3$. Then dynamic pressure is $p_w$ = 245 N/m$^2$. That is four time less than internal pressure $p$ = 1000 N/m$^2$. When the need arises, sometimes the internal pressure can be voluntarily decreased, bled off.

In Fig. 7 the alert reader has noticed: the daily heat loss is about the solar heat in the very coldest Winter day when a dome located above 60$^0$ North or South Latitude and the outside air temperature is –50 $^0$C. Let us compute and compare the heat extension for conventional buildings located on same region with and without the "Evergreen" dome.

Assume the two-story building, perhaps a home or small office, occupies 0.1 part of domed area. That means their walls and roof is equal to 0.5 part of dome area. Assume the building walls have a thickness of 0.2 m and are formed of baked bricks ($\lambda$ = 0.758 W/m·K). The differences of temperature are 20 + 50 = 70 $^0$C ($T_1$ = 293 $^0$K, $T_2$ = 223 $^0$K). So, 1 m$^2$ of building surface has a heat loss, in 1 second

$$q = 0.758 \times 0.5 \times 70/0.2 = 132.65\ \ W/m^2.s\,.$$

The radiation heat loss from 1 m$^2$ dome at night when the jalousies are closed tightly is

[Eq. (8-9)]:

$$C'_r = \frac{C_s}{1/\varepsilon_1 + 1/\varepsilon_2 - 1} = \frac{5.67}{1/0.05 + 1/0.9 - 1} = 0.28, $$
$$q = 0.5 C'_r[(0.01 T_1)^4 - (0.01 T_2)^4] = 6.86\ \ W/m^2 s \tag{14}$$

Transfer and convective heat losses is (for $\delta_1 = \delta_3 = 0.0001$ m, $\delta_2 = 1$ m, $\lambda_1 = \lambda_3 = 0.744$, $\lambda_2 = 0.0244$, $\alpha = 30$) (see Eq. (7)):

$$k = \frac{1}{2/30 + 0.0002/0.744 + 1/0.0244} = 0.025,$$
$$q = 0.5\pi k(t_1 - t_2) = 1.57 \times 0.025 \times 70 = 2.75\ \ W/m^2 \tag{15}$$

Transfer heat loss is 6.86 + 2.75 = 9.61 W/m$^2$ <132.62 W/m$^2$. The heat loss of the dome is less than 14 times the heat of the buildings inside unprotected by an inflated dome.

We consider a two-layer dome film and one heat screen. If needed, better protection can further reduce the head losses as we can utilize inflated dome covers with more layers and more heat screens. One heat screen decreases heat losses by 2, two screens can decrease heat flow by 3 times, three by 4 times, and so on. If the Polar Region domes have a mesh structure, the heat transfer decreases proportional to the summary thickness of its enveloping film layers.

**VI. MACROPROJECTS**

The EPZD innovations outlined here can be practically applied to other climatic regimes (from Polar to Tropical). We suggest initial macroprojects could be small (10 m diameter) houses (Fig. 3) followed by an "Evergreen" dome in the Arctic or Antarctica covering a land area 200 × 1000 m, with irrigated vegetation, homes, open-air swimming pools, playground.

The house and "Evergreen" dome have several innovations: Sun reflector, double transparent insulating film, controllable jalousies coated with reflective aluminum and an electronic cable mesh inherent to the film for dome safety/integrity monitoring purposes. By undertaking to construct a half-sphere house, we can acquire experience in such constructions and explore more complex constructions. By computation, a 10 m diameter home has a useful floor area of 78.5 $m^2$, airy interior volume of 262 $m^3$ covered by an envelope with an exterior area of 157 $m^2$. It film enclosure material would have a thickness of 0.0002 mm with a total mass of 65 kg.

A city-enclosing "Evergreen" dome of 200 × 1000 m (Fig. 1, with spherical end caps) could have calculated characteristics: useful area = 2.3 × $10^5$ $m^2$, useful volume 17.8 × $10^6$ $m^3$, exterior dome area of 3.75 × $10^5$ $m^2$ comprised of a film of 0.0002 mm thickness and 145 tonnes. If the "Evergreen" dome were formed with concrete 0.2 m thick, the mass of the city-size envelope would be 173 × $10^3$ tonnes, which is a thousand times heavier. Also, just for comparison, if we made a gigantic "Evergreen" dome with stiff glass, thousands of tonnes of steel, glass would be necessary and such materials would be very costly to transport hundreds, possibly thousands, of kilometers to the site where they would be assembled by highly-paid workers. Our non-woven textile (film) is flexible and plastic can be relatively cheap. The single greatest boon to "Evergreen" dome construction, whether in the Earth's Polar Regions or elsewhere, is the protected cultivation of plants within a dome that generates energy from the available and technically harnessed sunlight. However, the North and South Poles may, during the 21st Century, become places of cargo and passenger congregation since a Cable Space Transportation System, installed on Antarctica's ice-cap and on a floating artificial ice island has been proposed the would transfer people and things to and from the Moon.[19]

**VII. DISCUSSION**

As with any innovative macroproject proposal, the reader will naturally have many questions. We offer brief answers to the four most obvious questions our readers are likely to ponder.

(1) *How can snow and ice be removed from the dome?*

After a snowfall, weather conditions permitting, a helicopter can hover over the dome, blowing off the accumulated loose snow. Compacted snow and ice can be removed by activating remotely operated sprayers that squirt warmed anti-freeze liquid onto the dome's exterior. Such deicer must not be toxic like those currently used at airports.[20] The film cover is flexible and has a lift force of about 100 kg/$m^2$. We imagine that a controlled change of interior air pressure will serve to shake the snow and ice off. Such a technology is used in aircraft for wing de-icing.

(2) *Storm wind.*

This was thoroughly considered in Section V, above.

(3) *Cover damage.*

The envelope contains a cable mesh so that the film cannot be damaged greatly. Its double layering structure governs the escape of heated air inside the living zone. Electronic signals alert supervising personnel of any rupture problems and permit a speedy repair effort by well-trained persons.

(4) *What is the design life of the film covering?*

Depending on the kind of materials used, it may be as much a decade. In all or in part, the cover can be replaced periodically.

## VIII. CONCLUSION

"Evergreen" domes can foster the fuller economic development of cold regions such as the Earth's Arctic and Antarctic and, thus, increase the effective area of territory dominated by humans. Normal human health can be maintained by ingestion of locally grown fresh vegetables and healthful "outdoor" exercise. "Evergreen" domes can also be used in the Tropics and Temperate Zone. Eventually, they may find application on the Moon or Mars since a vertical variant, inflatable towers to outer space[21], are soon to become available for launching spacecraft inexpensively into Earth-orbit or interplanetary flights[22-26].

## REFERENCES

[1] Hall, R.E. and Jones, C.I. (1999) "Why Do Some Countries Produce So Much More Output Per Worker Than Others?", *Quarterly Journal Economics* 114: 83-116.

[2] Klein, S. and Halzen, F. (May 2005) "The ice cube at the end of the world", CERN Courier, pages 17-22.

[3] Hornig, J.F. (Ed.) *Social and Environmental Impacts of the James Bay Hydroelectric Project*, McGill-Queens University Press, Canada, 1999.

[4] Talley, I. (11 July 2006), "Arctic Harshness Hinders Search for Oil", *The Wall Street Journal* CCXLVIII: A10.

[5] Bolonkin, A.A. and R.B. Cathcart (2006), "Inflatable 'Evergreen' dome settlements for Earth's Polar Regions", *Clean Technologies and Environmental Policy* DOI 10.1007/s10098-006-0073-4.

[6] Floren, D.W., (2001), "Antarctic Mining Regimes: An Appreciation of the Attainable", *Journal of Environmental Law and Litigation* 16: 467.

[7] Fyfe, J.C. and Saenko, O.A. (2005), "Human-Induced Change in the Antarctic Circumpolar Current", *Journal of Climate* 18: 3068-3073.

[8] Overpeck, J.T. et al. (2005), "Arctic System on Trajectory to New, Seasonally Ice-Free State", *EOS, Transactions of the American Geophysical Union* 34: 309, 312-313.

[9] Walsh, Don (2007), "Losing its Cool", *Proceedings of the U.S. Naval Institute* 133: 88.

[10] Zhou, S. and Flynn, P.C. (2005), "Geoengineering Downwelling Ocean Currents: A Cost Assessment", *Climatic Change* 71: 203-220.

[11] Albright, L.D., (2001) "Environmental Control for Plants on Earth and in Space", *IEEE Control Systems Magazine* 21: 28-47.

[12] GOTO: http://www.edenproject.org.uk/

[13] Keating, W.R. and Donaldson, G.C. (2004), "The impact of global warming on health and mortality", *Southern Medical Journal* 97: 1093-1099.

[14] Hix, J., *The Glass House*. MIT Press, Cambridge, USA, 1974, pages 192-196.

[15] Braddock-Clarke, S.E. and M. O'Mahony, *Techno Textiles 2*. Thames & Hudson, NY, USA, 2006.

[16] McQuaid, M. (Ed.), *Extreme Textiles: Designing for High Performance*. Princeton Architectural Press, NY, USA, 2005.

[17] Koch, K-M. (Ed.), *Membrane Structures: Innovative Building with Film and Fabric*. Prestel, NY, USA. 2004.

[18] Naschekin, V.V., *Technical thermodynamic and heat transmission*. Public House High University, Moscow. 1969 (in Russian).

[19] Bolonkin, A.A. and R.B. Cathcart, (2006) "A Cable Space Transportation System at the Earth's Poles to Support Exploitation of the Moon", *Journal of the British Interplanetary Society* 59: 375-380.

[20] `Corsi, S.R. et al., (2006), "Characterization of Aircraft Deicer and Anti-Icer Components and Toxicity in Airport Snowbanks and Snowmelt Runoff", *Environ. Sci. Technol.* 40: 3195-3202.

[21] Bolonkin, A.A., (2003), "Optimal Inflatable Space Towers with 3-100 km Height", *Journal of the British Interplanetary Society* Vol. 56, pp. 87 - 97.

[22] Bolonkin A.A., (2006), "*Non-Rocket Space Launch and Flight*", Elsevier, London, 488 pgs.

[23] Bolonkin A.A., (2006), Cheap Textile Dam Protection of Seaport Cities against Hurricane Storm Surge Waves, Tsunamis, and Other Weather-Related Floods. Printed in http://arxiv.org .

[24] Bolonkin A.A., (2006), Control of Regional and Global Weather. Printed in http://arxiv.org .

[25] Bolonkin, A.A. and R.B. Cathcart (2006), Antarctica: A Southern Hemisphere Windpower Station? Printed in http://arxiv.org .

[26] Bolonkin, A.A. and R.B. Cathcart (2006), A Low-Cost Natural Gas/Freshwater Aerial Pipeline. Printed in http://arxiv.org .